\documentclass[10pt,aps,prd,nofootinbib,superscriptaddress,twocolumn,preprintnumbers,balancelastpage,reprint]{revtex4-2}

\usepackage{graphicx,epsfig,psfrag,amssymb,hyperref}
\usepackage{multirow}
\usepackage{color,graphicx,epsfig,psfrag,amsmath,empheq}
\usepackage{bm}
\usepackage{mathrsfs,amsfonts,color}
\usepackage{slashed}
\usepackage{hepunits}
\usepackage{color}
\usepackage{enumitem}
\usepackage{xspace}
\usepackage{comment}

\newcommand{\gluex}{\textsc{GlueX}\xspace}

\def\diphoton {\ensuremath{\gamma\gamma}\xspace}
\def\threepi {\ensuremath{\pi^+\pi^-\pi^0}\xspace}

\begin{document}
\title{Search for photoproduction of axion-like particles at GlueX
}

\affiliation{Arizona State University, Tempe, Arizona 85287, USA}
\affiliation{National and Kapodistrian University of Athens, 15771 Athens, Greece}
\affiliation{Carnegie Mellon University, Pittsburgh, Pennsylvania 15213, USA}
\affiliation{The Catholic University of America, Washington, D.C. 20064, USA}
\affiliation{University of Connecticut, Storrs, Connecticut 06269, USA}
\affiliation{Duke University, Durham, North Carolina 27708, USA}
\affiliation{Florida International University, Miami, Florida 33199, USA}
\affiliation{Florida State University, Tallahassee, Florida 32306, USA}
\affiliation{The George Washington University, Washington, D.C. 20052, USA}
\affiliation{University of Glasgow, Glasgow G12 8QQ, United Kingdom}
\affiliation{GSI Helmholtzzentrum f\"{u}r Schwerionenforschung GmbH, D-64291 Darmstadt, Germany}
\affiliation{Institute of High Energy Physics, Beijing 100049, People's Republic of China}
\affiliation{Indiana University, Bloomington, Indiana 47405, USA}
\affiliation{Alikhanov Institute for Theoretical and Experimental Physics NRC Kurchatov Institute, Moscow 117218, Russia}
\affiliation{IKP, Forschungszentrum J\"{u}lich, D-52428 J\"{u}lich GmbH, Germany}
\affiliation{Lamar University, Beaumont, Texas 77710, USA}
\affiliation{University of Massachusetts, Amherst, Massachusetts 01003, USA}
\affiliation{Massachusetts Institute of Technology, Cambridge, Massachusetts 02139, USA}
\affiliation{National Research Nuclear University Moscow Engineering Physics Institute, Moscow 115409, Russia}
\affiliation{Norfolk State University, Norfolk, Virginia 23504, USA}
\affiliation{North Carolina A\&T State University, Greensboro, North Carolina 27411, USA}
\affiliation{University of North Carolina at Wilmington, Wilmington, North Carolina 28403, USA}
\affiliation{Old Dominion University, Norfolk, Virginia 23529, USA}
\affiliation{University of Regina, Regina, Saskatchewan S4S 0A2, Canada}
\affiliation{Universidad T\'ecnica Federico Santa Mar\'ia, Casilla 110-V Valpara\'iso, Chile}
\affiliation{Thomas Jefferson National Accelerator Facility, Newport News, Virginia 23606, USA}
\affiliation{Tomsk State University, 634050 Tomsk, Russia; Tomsk Polytechnic University, 634050 Tomsk, Russia}
\affiliation{Union College, Schenectady, New York 12308, USA}
\affiliation{Washington \& Jefferson College, Washington, Pennsylvania 15301, USA}
\affiliation{William \& Mary, Williamsburg, Virginia 23185, USA}
\affiliation{Wuhan University, Wuhan, Hubei 430072, People's Republic of China}
\affiliation{A. I. Alikhanian National Science Laboratory (Yerevan Physics Institute), 0036 Yerevan, Armenia}
\author{S.~Adhikari} \affiliation{Old Dominion University, Norfolk, Virginia 23529, USA}
\author{C.~S.~Akondi} \affiliation{Florida State University, Tallahassee, Florida 32306, USA}
\author{M.~Albrecht} \affiliation{Indiana University, Bloomington, Indiana 47405, USA}
\author{A.~Ali} \affiliation{GSI Helmholtzzentrum f\"{u}r Schwerionenforschung GmbH, D-64291 Darmstadt, Germany}
\author{M.~Amaryan} \affiliation{Old Dominion University, Norfolk, Virginia 23529, USA}
\author{A.~Asaturyan} \affiliation{A. I. Alikhanian National Science Laboratory (Yerevan Physics Institute), 0036 Yerevan, Armenia}
\author{A.~Austregesilo} \affiliation{Thomas Jefferson National Accelerator Facility, Newport News, Virginia 23606, USA}
\author{Z.~Baldwin} \affiliation{Carnegie Mellon University, Pittsburgh, Pennsylvania 15213, USA}
\author{F.~Barbosa} \affiliation{Thomas Jefferson National Accelerator Facility, Newport News, Virginia 23606, USA}
\author{J.~Barlow} \affiliation{Florida State University, Tallahassee, Florida 32306, USA}
% ...
\collaboration{The \textsc{GlueX} Collaboration}

\begin{abstract}
 We present a search for axion-like particles, $a$, produced in photon-proton collisions at a center-of-mass energy of approximately 4\,GeV, focusing on the scenario where the $a$-gluon coupling is dominant.
 The search uses $a\to\gamma\gamma$ and $a\to\pi^+\pi^-\pi^0$ decays, and a data sample corresponding to an integrated luminosity of 168 pb$^{-1}$ collected with the \gluex detector. The search for $a\to\gamma\gamma$ decays is performed in the mass range of $180 < m_a < 480$\,MeV, while the search for $a\to\pi^+\pi^-\pi^0$ decays
  explores the $600 < m_a < 720$\,MeV region.
  No evidence for a signal is found,
  and 90\% confidence-level exclusion limits are placed on the $a$-gluon coupling strength. These constraints are the most stringent to date over much of the mass ranges considered.
\end{abstract}

\maketitle

\section{Introduction}

Axion-like particles (ALPs), $a$, are hypothetical pseudoscalars found in many proposed extensions to the Standard Model (SM)~\cite{Essig_2013, Marsh_2016, Graham_experiment_2015, Irastorza_2018}.
ALPs can naturally address the Strong CP~\cite{Peccei_cp_1977, Peccei_constraints_1977, Weinberg_1978, Wilczek_1978} and Hierarchy problems~\cite{Graham_hierarchy_2015},
and could provide a \textit{portal} that connects SM particles to dark matter~\cite{Nomura_portal_2009,Freytsis:2010ne,Dolan:2014ska,Hochberg:2018rjs}.
The couplings of ALPs to the SM are highly suppressed at low energies by a large cutoff scale $\Lambda$; however, since ALPs are pseudo-Nambu-Goldstone bosons, their mass $m_a$ can be much smaller than the scale that controls their dynamics, {\em i.e.}, $m_a\ll\Lambda$.

Recently, ALPs with $\Lambda_{\rm QCD}$-scale masses whose dominant coupling to the SM is with the gluonic field have received considerable interest~\cite{Marciano:2016yhf,Jaeckel:2015jla,Dobrich:2015jyk,Izaguirre:2016dfi,Knapen:2016moh,Williams_ALP_pheno, Williams_ALP_photoproduction,Kelly:2020dda}.
Such $m_a$ values can be obtained, {\em e.g.}, by introducing a mirror strongly-coupled sector that generates a large ALP potential which aligns with the QCD-generated axion potential~\cite{Hook:2014cda,Fukuda:2015ana,Dimopoulos:2016lvn,Hook:2019qoh}.
In this scenario, the ALP solves the Strong~CP problem, and because $m_a \gtrsim \Lambda_{\rm QCD}$, it is robust against UV contributions that would otherwise give rise to the well-known Quality problem~\cite{Kamionkowski:1992mf,Barr:1992qq,Ghigna:1992iv,Holman:1992us}.

The effective Lagrangian
describing ALP-gluon interactions is
\begin{equation}
  \mathcal{L}\supset-\frac{4\pi\alpha_s c_g}{\Lambda}aG^{\mu\nu}\tilde{G}_{\mu\nu},
\end{equation}
where $c_g$ is the dimensionless $agg$ vertex coupling constant and $\tilde{G}_{\mu\nu}\equiv\frac{1}{2}\epsilon_{\mu\nu\alpha\beta}G^{\alpha\beta}$.
Reference~\cite{Williams_ALP_pheno} presented a novel data-driven method for
determining the hadronic interaction strengths of such ALPs, and showed that the dominant decay modes for $m_a \lesssim 0.8$\,GeV ($c=1$ throughout this article) are to the \diphoton, $3\pi$, and $\pi^+\pi^-\gamma$ final states.
In a follow-up article~\cite{Williams_ALP_photoproduction}, photoproduction of
ALPs was explored, including the discovery potential of the \gluex experiment.
In addition, Ref.~\cite{Williams_ALP_photoproduction} presented a search based on the published \gluex result~\cite{GlueX_2017} on $\gamma p \to p \gamma\gamma$ that set world-leading limits in some regions
of the ALP parameter space.

In this article,  we present a search for ALPs produced in photon-proton collisions at a center-of-mass energy $\sqrt{s} \approx 4$\,GeV, focusing on the scenario where the $a$-gluon coupling is dominant.
The search uses $a\to\gamma\gamma$ and $a\to\pi^+\pi^-\pi^0$ decays, and a data sample corresponding to an integrated luminosity of 168 pb$^{-1}$ collected with the \gluex detector in the beam-energy range $8 < E_{\textrm{beam}} < 9$\,GeV.
The search for $a\to\gamma\gamma$ decays is performed in the mass range of $180 < m_a < 480$\,MeV, while the search for $a\to\pi^+\pi^-\pi^0$ decays
explores the $600 < m_a < 720$\,MeV region.
These mass ranges are chosen to avoid the narrow peaks from the $\pi^0$, $\eta$, and $\omega$ mesons, and focus on where {\sc GlueX} is expected to have world-leading sensitivity.
Only the charged $3\pi$ ALP final state is considered, since the $a \to 3 \pi^0$ decay is more challenging to study using the \gluex detector.
Similarly, the other dominant decay mode predicted by Ref.~\cite{Williams_ALP_pheno} at low masses, namely $a \to \pi^+\pi^-\gamma$, is not considered due to the huge background from final states such as $\gamma p\to p \pi^+\pi^-\pi^0$ with an undetected photon, or $\gamma p \to p\pi^+\pi^-$ with an additional photon from hadronic split-off interactions of the charged pions in the calorimeter.

Following Ref.~\cite{Williams_ALP_pheno}, we denote the $m_a$-dependent mixing terms between the ALP and pseudoscalar mesons as
$\langle\boldsymbol{a\pi^0}\rangle$ and $\langle\boldsymbol{a\eta}\rangle$.
The expected ALP yield in final state $\mathcal{F}$ in a small bin of $[s, t]$, where $t\equiv (p_a - p_{\rm beam})^2$, is related to the observed $\pi^0 \to \mathcal{F}$ and $\eta \to \mathcal{F}$ yields in the bin, $n_{\pi^0}(s,t)$ and $n_{\eta}(s,t)$,  by~\cite{Williams_ALP_photoproduction}
\begin{equation}
  \begin{split}
    n_a(s,t) &\approx  \left(\frac{f_\pi}{f_a}\right)^2 \bigg[|\langle\boldsymbol{a\pi^0}\rangle|^2\frac{n_{\pi^0}(s,t)\epsilon(m_a,s,t)}{\mathcal{B}(\pi^0\to\mathcal{F})\epsilon(m_\pi,s,t)} \\& +|\langle\boldsymbol{a\eta}\rangle|^2\frac{n_{\eta}(s,t)\epsilon(m_a,s,t)}{\mathcal{B}(\eta\to\mathcal{F})\epsilon(m_\eta,s,t)}\bigg]\mathcal{B}(a\to\mathcal{F}),
  \end{split}
  \label{eq:norm}
\end{equation}
where
$f_\pi$ and $f_a\equiv-\Lambda/32\pi^2c_g$ are the pion and ALP decay constants,
$\mathcal{B}(\pi^0,\eta\to\mathcal{F})$ are the
known meson-decay branching fractions~\cite{PDG},
$\epsilon$ denotes the $m_a$-dependent product of the detector acceptance and efficiency,
and $\mathcal{B}(a\to\mathcal{F})$ is the  ALP-decay branching fraction~\cite{Williams_ALP_pheno}.
Equation~\eqref{eq:norm} assumes that $t$-channel processes are dominant, which is known to be true at \gluex energies for $-t \lesssim 1$\,GeV$^2$\,\cite{GlueX_2017}.
For the \threepi decay, the ALP-pion mixing term is negligible and is ignored.

The use of Eq.~\eqref{eq:norm} facilitates a largely data-driven search where most experimental systematic uncertainties cancel.
For example, knowledge of the luminosity and absolute efficiencies is not required; only the relative efficiency to reconstruct the ALP and pseudoscalar-meson decays to the same final state is needed.
The fiducial regions used, defined in Table~\ref{tab:fid}, ensure that the detector response is sufficiently model independent and that $t$-channel production processes are dominant.
This not only reduces the systematic uncertainties that contribute to the results presented here, but enables recasting our results for other models, for which
sufficient information has been provided as Supplemental Material~\cite{Supp}.

\begin{table}[t]
\centering
\caption{Fiducial regions of the searches for both $a \to \diphoton$ and $a \to \threepi$ decays, defined in terms of the photon beam energy, $E_\text{beam}$, the momentum of the outgoing proton, $p_p$, Mandelstam $t$, the polar angle of the outgoing photons, $\theta_\gamma$, the opening angle between the outgoing photon pair, $\alpha(\gamma_1,\gamma_2)$, and the energy of the outgoing photons, $E_\gamma$. }
\label{tab:fid}
\begin{tabular}{cc}
\hline
\centering
\multirow{5}{*}{All searches}
& $8<E_{\text{beam}}<9\,\text{GeV}$ \\
& $p_p>0.35\,\text{GeV}$ \\
& $-t<1\,\text{GeV}^2$ \\
& $2.45<\theta_\gamma<9.7^\circ$, $11.66<\theta_\gamma<37.4^\circ$ \\
& $\alpha(\gamma_1,\gamma_2)>1.15^\circ$ \\
\hline
\multirow{2}{*}{$\gamma\gamma$ channel}
& $0.5<E_\gamma<10\,\text{GeV}$ \\
& $-t > 0.2$\,GeV$^2$ \\
\hline
\multirow{2}{*}{$\pi^+\pi^-\pi^0$ channel}
& $0.1<E_\gamma<10\,\text{GeV}$ \\
& $-t > 0.15$\,GeV$^2$ \\
\hline
\end{tabular}
\end{table}

\section{Experiment \& Simulation}

This search is performed using the \gluex spectrometer located in Hall~D at Jefferson Lab.
A tagged linearly polarized photon beam is created from the 11.6\,GeV electron beam at the Continuous Electron Beam Accelerator Facility~\cite{CEBAF2001} by coherent bremsstrahlung off a diamond radiator.
This search does not use the polarization information.
The momenta of the scattered beam electrons are measured using a scintillating-fiber array, which enables measuring the photon-beam energy to about 0.1\% precision in the energy region used in this search.
The photon beam passes through a collimator that highly suppresses its incoherent component, before reaching the liquid hydrogen target.
This collimation results in most of the high-energy flux being in the small energy range considered here, which motivates our choice of using only a single $s$ bin.

The \gluex spectrometer~\cite{GlueX_NIM}, which has a nearly hermetic angular coverage, consists of the liquid hydrogen target surrounded by a scintillator start counter~\cite{SC_NIM}, a straw-tube central drift chamber~\cite{CDC_NIM}, and a lead and scintillating-fiber barrel calorimeter~\cite{BCAL_NIM}, all of which are inside the bore of a superconducting solenoid.
In the forward region downstream of the central drift chamber, there are
four sets of planar wire drift chambers~\cite{FDC_NIM} inside of the solenoid, with a time-of-flight scintillator wall and a forward lead-glass calorimeter~\cite{FCAL_NIM} located further downstream and outside of the solenoid.
The trigger selects events with sufficient energy deposited in the calorimeters, while the event selection for the reactions of interest is performed offline.
The drift chambers provide momentum and energy loss measurements for charged particles.
The calorimeters measure the energy and position of electromagnetic showers induced by both charged and neutral particles.
Finally, particle identification and beam-photon selection are based on time-of-flight measurements obtained using information from the start counter, the calorimeters, and the time-of-flight wall.

Simulation is required to model the effects of the \gluex detector acceptance and its response to the reactions studied here.
Specifically, the relative efficiency to reconstruct the ALP and pseudoscalar-meson decays to the same final state is required in Eq.~\eqref{eq:norm}.
In the simulation, the reactions are generated using the \textsc{Genr8} package~\cite{genr8}, which assumes purely $t$-channel production.
The effects of the interactions of other beam photons, including accidental coincidences and random detector backgrounds, are included in the simulation.
The interaction of the generated particles with the \gluex detector, and its response, are implemented using the \textsc{Geant4} toolkit~\cite{geant4}.

\section{Event Selection}
\label{sec:selection}

The search is based on events containing candidates for the exclusive reactions $\gamma p \to p \diphoton$ or $\gamma p \to p \threepi$, with the subsequent decay $\pi^0\to\diphoton$.
Candidate reactions must satisfy all of the fiducial requirements in Table~\ref{tab:fid}.
In order to avoid experimenter bias, all aspects of the selection are fixed without examining the evidence for an ALP signal.

Events must have the exact number of positively and negatively charged tracks required for each reaction, {\em i.e.}\ events with additional tracks are discarded.
Events must also have at least one tagged beam photon candidate, and at least two neutral shower candidates.
Additional tagged beam photons and showers are permitted in the initial selection; however, the total energy of the latter is required to be less than 100\,MeV.

Charged particles, namely protons and charged pions, are identified using time-of-flight information, along with
the energy loss of their tracks in the drift chambers.
The absolute value of the squared missing mass for each reaction is required to be less than 0.05\,GeV$^2$.
For $a \to \threepi$ candidates, the diphoton invariant mass must be consistent with that of the $\pi^0$ meson.
A kinematic fit is used to select particle combinations that are consistent with conservation of energy and momentum, and subsequently, to improve the experimental resolution by enforcing these conservation laws (see Ref.~\cite{GlueX_NIM} for details).
Finally, the photon-proton collision must be consistent with having occurred within the liquid hydrogen target, as determined using the tracking information from the final-state charged particles.

The use of time-of-flight information to perform charged-particle identification enforces that
beam-photon candidates have an arrival time,
based on the precise electron-beam timing information,
at the photon-proton collision point that is consistent with when the final-state particles were produced, obtained using information recorded by the \gluex spectrometer.
However, incorrect beam-photon candidates that satisfy both the timing and kinematic requirements do occur, typically due to additional photons in the same beam bunch with similar energies.
This accidental background is statistically subtracted using a data sample of out-of-time beam photons.

\begin{figure}[tb]
  \centering
  \includegraphics[width=0.49\textwidth]{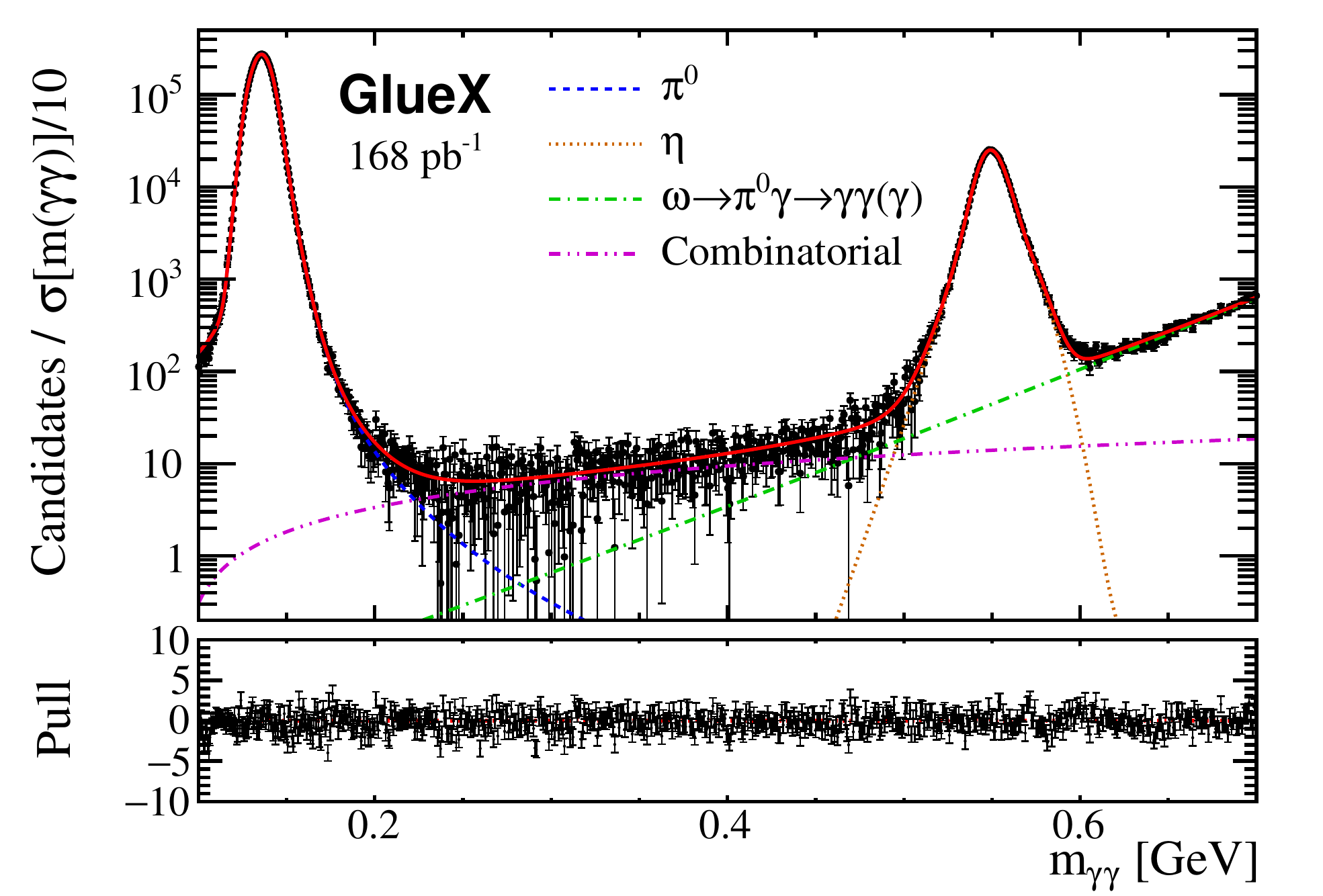}
  \includegraphics[width=0.49\textwidth]{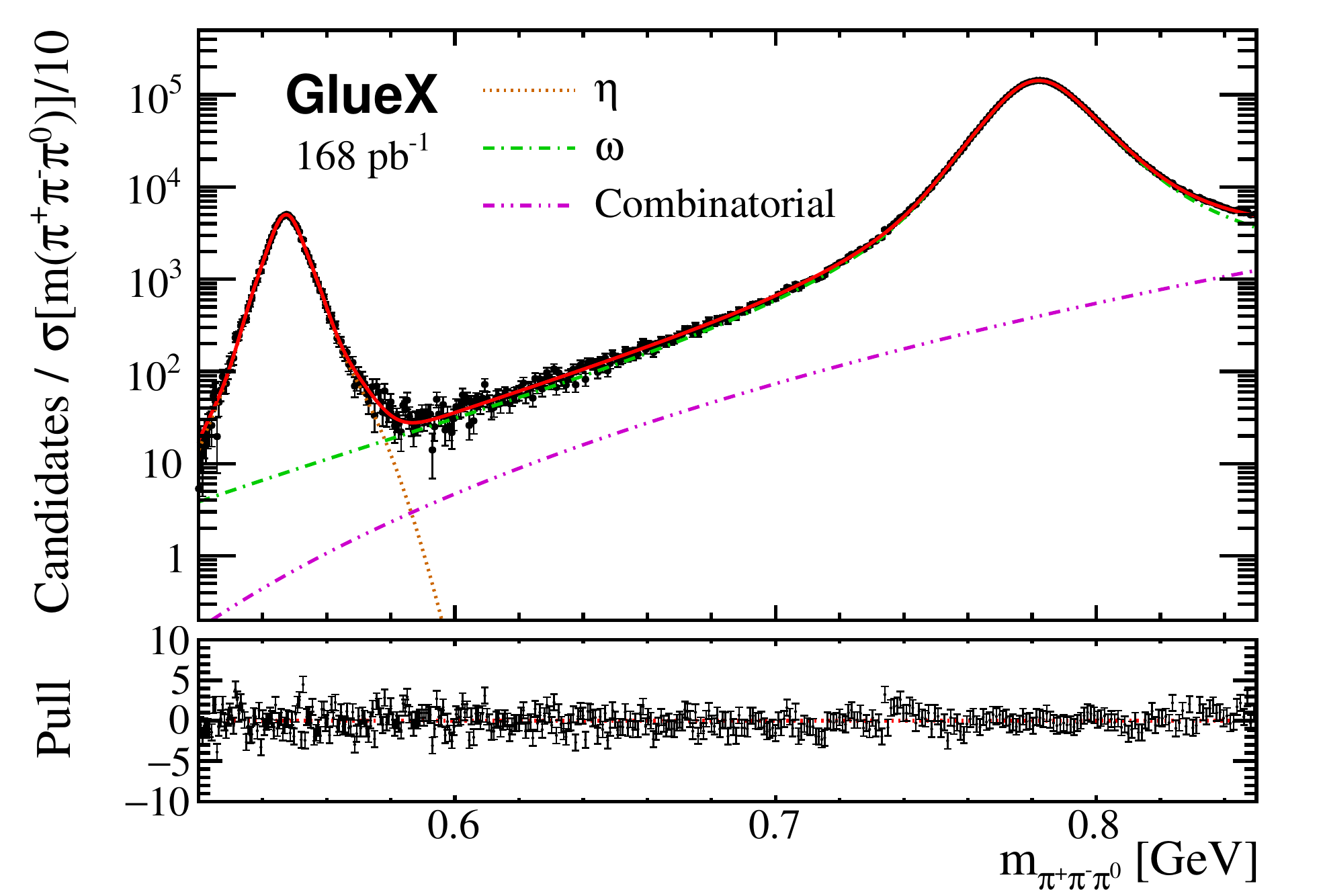}
  \caption{Fits to the (top) \diphoton and (bottom) \threepi invariant mass spectra after applying the accidental subtraction used to determine the $\pi^0$ and $\eta$ yields (shown here integrated over $t$). The blue dashed line shows the $\pi^0$ component, the orange dotted lines show the $\eta$ components, the green dashed-dotted lines show the $\omega$ components, and the magenta dashed-double-dotted line shows the linear (power-law) background for the \diphoton (\threepi) channel. The red solid lines show the sum of all contributions. The pulls account for both the statistical and modeling uncertainties. The mass-dependent variable bin size is determined according to the mass resolution function.
  }
  \label{fig:meson_mass_fits}
\end{figure}

Figure~\ref{fig:meson_mass_fits} shows the \diphoton and \threepi invariant mass spectra obtained after applying the full selection.
Fits are performed to these spectra in bins of $t$ to obtain the observed $\pi^0$ and $\eta$ yields needed in Eq.~\eqref{eq:norm}; Fig.~\ref{fig:meson_mass_fits} shows the fit results integrated over $t$.
It is worth noting that the bump-hunt procedure we use to search for ALP signals, described in detail in the next section, introduces additional complexity to the background models beyond what is used here.
The fits shown in Fig.~\ref{fig:meson_mass_fits} are only used to determine the meson yields and the mass resolution.

For the \diphoton channel, the fit model consists of the following components.
The $\pi^0$ and $\eta$ components are modeled by sums of Gaussian and Crystal Ball~\cite{Skwarnicki:1986xj} functions with power-law tails on both sides of the peaks.
The $\omega \to \pi^0\gamma \to \diphoton(\gamma)$ component, where one of the three photons is not detected, is modeled by a sum of two Crystal Ball functions with power-law tails on opposite sides of the peak.
A combinatorial background component is modeled by a linear function.
The $\pi^0$ and $\eta$ yields are $4.4\pm0.1$ million and $0.62\pm0.02$ million, respectively.
Here, the systematic uncertainties, which are obtained by varying both the pseudoscalar and background models, are dominant.
The variations include using different numbers of Gaussian and Crystal Ball functions for the pseudoscalar models and using different orders of the polynomial for the background model.
In addition, the \diphoton mass resolution, defined as half the width of the region containing 68\% of the signal probability, is determined to be about 6\,(9)\,MeV at $m_{\pi^0}\, (m_{\eta})$.
Simulation is used to interpolate between these values to obtain the resolution in the $m_{\diphoton}$ region considered in the ALP search with a precision of 2\%.

The components used in the fit model for the \threepi channel are as follows.
The $\eta$ component is modeled by a double Gaussian.
The known $\omega$ lineshape is taken from Ref.~\cite{omega_lineshape} and then convolved with a resolution function modeled by a sum of Crystal Ball functions with power-law tails on both sides of the $\omega$ peak.
A combinatorial background component is modeled by a power-law function of the energy released in the \threepi system.
The $\eta$ yield is found to be $70\pm1$ thousand, where the stated uncertainty is dominated by systematic errors.
The \threepi mass resolution is determined to be about 6\,(11)\,MeV at $m_{\eta}\,(m_{\omega})$.
Simulation is again used to interpolate between these values to obtain the resolution in the $m_{\threepi}$ region considered in the ALP search with 2\% precision.

\section{Signal Searches}
\label{sec:search}

The signal-search strategy and method, which were first introduced in Ref.~\cite{Williams_bump_hunt}, are similar to those used in Refs.~\cite{LHCb_dark_photon_2018,LHCb_dark_photon_2020,Aaij:2020ikh}.
The mass spectra for ALP final states $\mathcal{F} = \diphoton$ and $\threepi$ are scanned in steps of about half the mass resolution, $\sigma(m_{\mathcal{F}})/2$, searching for ALP contributions.
At each $m_a$ hypothesis, a binned extended maximum-likelihood fit is performed in a $\pm 12.5\sigma(m_{\diphoton})$ or $\pm 7.5 \sigma(m_{\threepi})$ window around $m_a$; a narrower window is used in the \threepi final state due to the small distance between the $\eta$ and $\omega$ peaks compared to $\sigma(m_{\threepi})$.
The profile likelihood method is used to determine the local $p$-values and the ALP signal-yield confidence intervals.
The trial factors are obtained using pseudoexperiments for each final state, {\em i.e.}\ by generating a large ensemble of background-only datasets, and rerunning the full signal-search procedure on each sample.
The \textit{bounded likelihood} approach~\cite{Rolke:2004mj} is used when determining the confidence intervals, which defines $\Delta \log{\mathcal{L}}$ relative to zero signal, instead of the best-fit value, if the best-fit signal value is negative.
This approach has two benefits:
it enforces that only physical (nonnegative) upper limits are placed on the ALP yields,
and it prevents these limits from being much better than the experimental sensitivity if a large deficit in the background yield is observed.

The ALP signal mass distributions are well modeled by a Gaussian function, whose resolution is determined precisely as described in the previous section.
The mass-resolution uncertainty is included in the profile likelihood.
A small correction is applied to remove the bias due to neglecting non-Gaussian components of the signal shape, which is determined to be about 1\% from the large $\pi^0$ and $\eta$ peaks observed in the data.

The background models include the meson components described in the previous section and shown in Fig.~\ref{fig:meson_mass_fits}.
In addition, Legendre polynomial terms up to $\ell = 4 \, (2)$ for $\mathcal{F} = \diphoton \, (\threepi)$ are taken as inputs, then the data-driven model-selection process of Ref.~\cite{Williams_bump_hunt} is performed.
This approach is necessary due to the unknown origins of much of the backgrounds.
The uncertainty of the model-selection process is included in the profile likelihood following Ref.~\cite{Dauncey:2014xga}.
Specifically, the \textit{aic-o} method in Ref.~\cite{Williams_bump_hunt} is used, which penalizes the log-likelihood of each background model according to its complexity (number of parameters).
The confidence intervals are obtained from the penalized profile likelihoods treating the model index as a discrete nuisance parameter~\cite{Dauncey:2014xga}.

The maximum $\ell$ values are chosen for each final state to ensure adequate description is possible for any peaking background that may contribute.
Where such complexity is unnecessary, the data-driven model-selection procedure effectively reduces the
complexity to increase the sensitivity because the associated penalty terms result in overly complex models being ignored when constructing the confidence intervals (see Ref.~\cite{Williams_bump_hunt} for more detailed discussion).
Following Ref.~\cite{Williams_bump_hunt}, all fit regions are
transformed onto the interval $[-1, 1]$ with the scan $m_a$ value mapped to zero.
The signal model is an even function after this transformation; therefore, the presence of odd
Legendre modes, which are orthogonal to the signal component, has minimal impact on the variance of the ALP yield.
Thus, all odd Legendre modes are included in every background model, while only a subset of the even modes is selected.
This procedure results in a mass-dependent background-model uncertainty, which on average is about 5\%.

\begin{figure}[tb]
  \centering
  \includegraphics[width=0.45\textwidth]{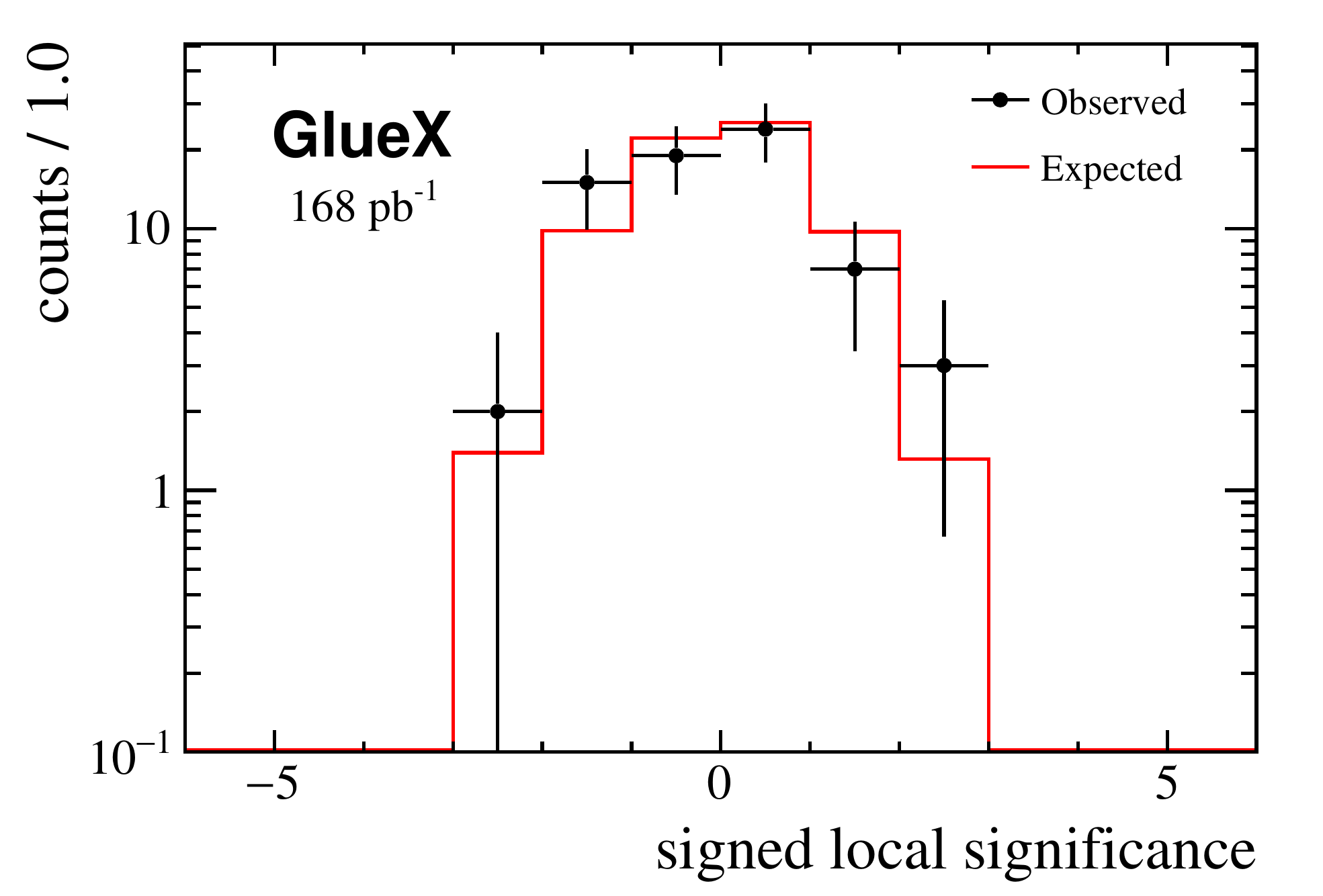}
  \includegraphics[width=0.45\textwidth]{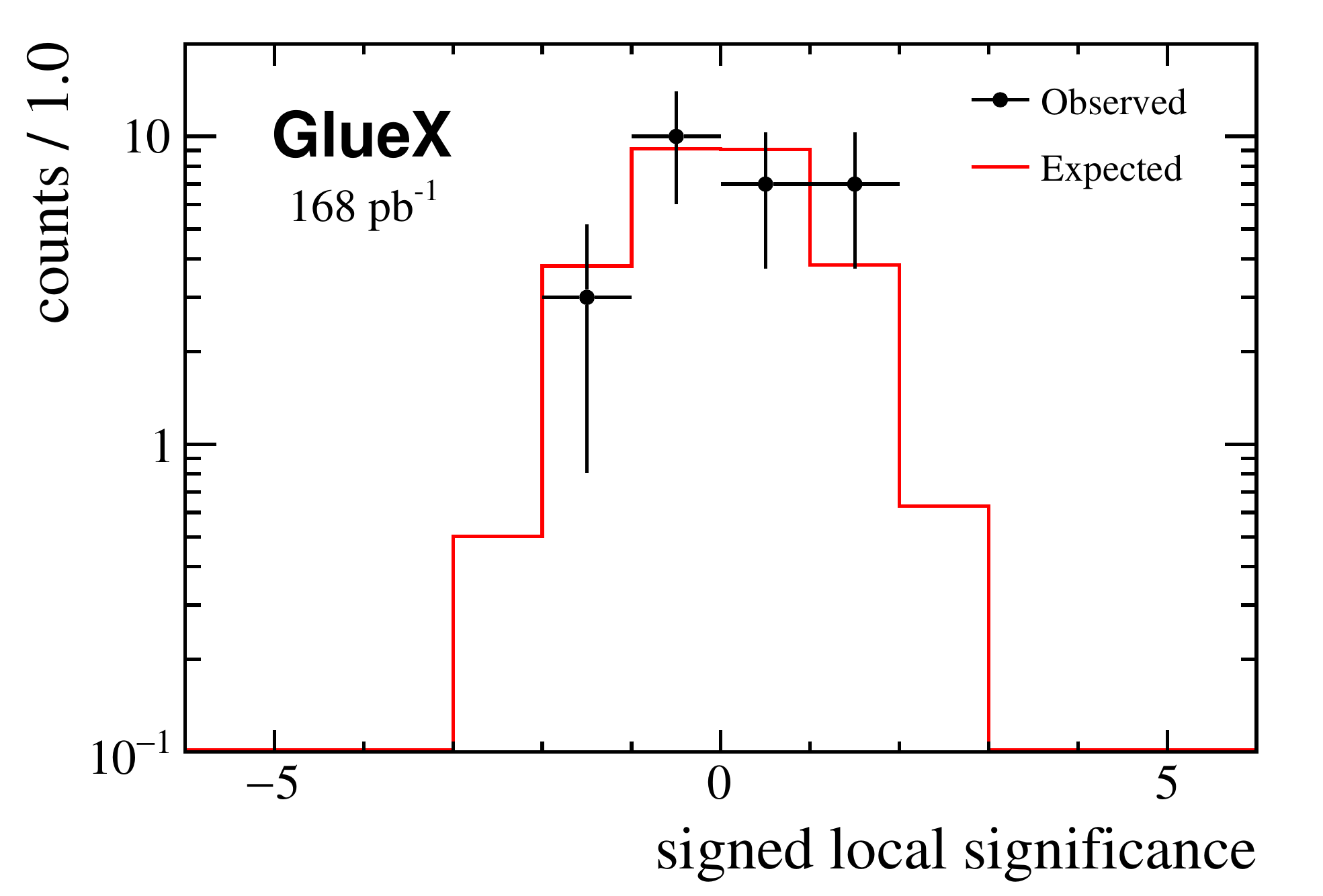}
  \caption{Signed local significances at each scan mass from (black points) all fits and (red lines) the expected distribution for the (top) \diphoton and (bottom) \threepi channels; if the best-fit signal-yield estimator is negative, the signed significance is negative and \textit{vice versa}. The error bars account for the correlation between nearest-neighbor fit results, which often produces outliers in pairs due to the $\sigma(m)/2$ step size.
  }
  \label{fig:significance}
\end{figure}

Figure~\ref{fig:significance} shows the signed local significances for all $m_a$ values scanned for both ALP decays considered.
The largest local excess in the \diphoton spectrum is $2.3\sigma$ at 213\,MeV.
Similarly, the largest local excess in the \threepi spectrum is $1.6\sigma$ at 669\,MeV. The global p-value is found to be 0.28\,(0.50) for the \diphoton\,(\threepi) channel, after accounting for the trials factor due to the number of signal hypotheses.

\section{Acceptance \& Efficiency}
\label{sec:efficiency}

The use of Eq.~\eqref{eq:norm} for normalization means that only the relative efficiency to reconstruct the ALP and pseudoscalar-meson decays to the same final state is needed.
Furthermore, since the normalization is done in narrow $[s,t]$ bins, the production mechanisms do not need to be well understood.
The acceptance is defined here as the probability that a reaction producing an ALP in a $[s,t]$ bin will have all final-state particles in the fiducial region defined in Table~\ref{tab:fid}.
This acceptance is strongly dependent on $m_a$ and requires careful treatment as described below.
For accepted reactions, the reconstruction efficiency has minimal dependence on $m_a$ or $t$ and is taken from simulation.
Indeed, our choice of fiducial region is designed to minimize the $m_a$ and $t$ dependence, since only this dependence enters into Eq.~\eqref{eq:norm}.
The uncertainty due to the relative reconstruction efficiency is, therefore, negligible compared to that of the acceptance.

The acceptance varies strongly with both $m_a$ and $t$.
The $t$ bins used for the normalization are only  0.05\,GeV$^2$ wide, which reduces the impact of the $t$ dependence, but it still must be accounted for.
To do this, we numerically sample from the phase-space for each $t$ bin and obtain the probability that the reaction satisfies the fiducial requirements.
In each $t$ bin, we consider three scenarios: $(i)$ $t$ fixed to the lower edge of the bin,
$(ii)$ $t$ fixed to the upper edge of the bin, and
$(iii)$ $t$ sampled uniformly over the bin range.
Since the bins are narrow, scenario $iii$ is used to obtain the nominal acceptance.
Half the difference of $i$ or $ii$ from nominal, whichever difference is larger is used, is taken as the systematic uncertainty in the acceptance in each $t$ bin.
Bins whose acceptance uncertainty is larger than 10\% are excluded from the fiducial region.

%Figure~\ref{fig:acc_eff} shows
The product of the acceptance and efficiency for each ALP decay as a function of $m_a$ and $t$ is provided in Supplemental Material.
The acceptance uncertainties in each bin, obtained as described in the previous paragraph, are propagated to the expected ALP yield using Eq.~\eqref{eq:norm} for each $m_a$ value considered in the search.
These uncertainties, which vary with $m_a$, are typically about 5\%.

\begin{comment}
\begin{figure}[t]
  \centering
  \includegraphics[width=0.45\textwidth]{figs/map_gg_acc_eff_tBin_mass.pdf}
  \includegraphics[width=0.45\textwidth]{figs/map_pi0pippim_acc_eff_tBin_mass.pdf}
  \caption{Products of the acceptance and efficiency in bins of $t$ and $m_a$ for (top) $a \to \diphoton $ and (bottom) $a \to \threepi$ decays.}
  \label{fig:acc_eff}
\end{figure}
\end{comment}

\section{Results}

The upper limits on the ALP signal yields obtained in Sec.~\ref{sec:search} are normalized using Eq.~\eqref{eq:norm}, which takes as input the pseudoscalar-meson yields from Sec.~\ref{sec:selection}, relative efficiency from Sec.~\ref{sec:efficiency},
the pseudoscalar-meson decay branching fractions from the PDG\,\cite{PDG},
and the ALP decay branching fractions from Ref.~\cite{Williams_ALP_pheno}.
The systematic uncertainties on the ALP signal yield, the pseudoscalar-meson yields and branching fractions, and on the relative efficiency are included in the profile likelihood when determining the upper limits on the ALP yield.
These uncertainties, which were described previously, are summarized in Table~\ref{tab:systematics}.

\begin{table}[tb]
\centering
\caption{Summary of relative systematic uncertainties. Those specified as a range are mass dependent.}
\label{tab:systematics}
\begin{tabular}{ccc}
\hline
Source                       &  $a \to \diphoton$ & $a \to \threepi$ \\ \hline
Signal model                 & 1\%  & 1\% \\
Background model             & 2--10\% & 2--8\%   \\ \hline
Acceptance$\times$efficiency & 3--6\% & 5\% \\
$\pi^0$ and $\eta$ yields    & 3\%   &  1\% \\
Branching fractions          & 0.1--0.5\%  & 1.5\%\\ \hline
Total                        & 5--12\%      & 5--9\%        \\
\hline
\end{tabular}
\end{table}

\begin{figure}[tb]
  \centering
  \includegraphics[width=0.45\textwidth]{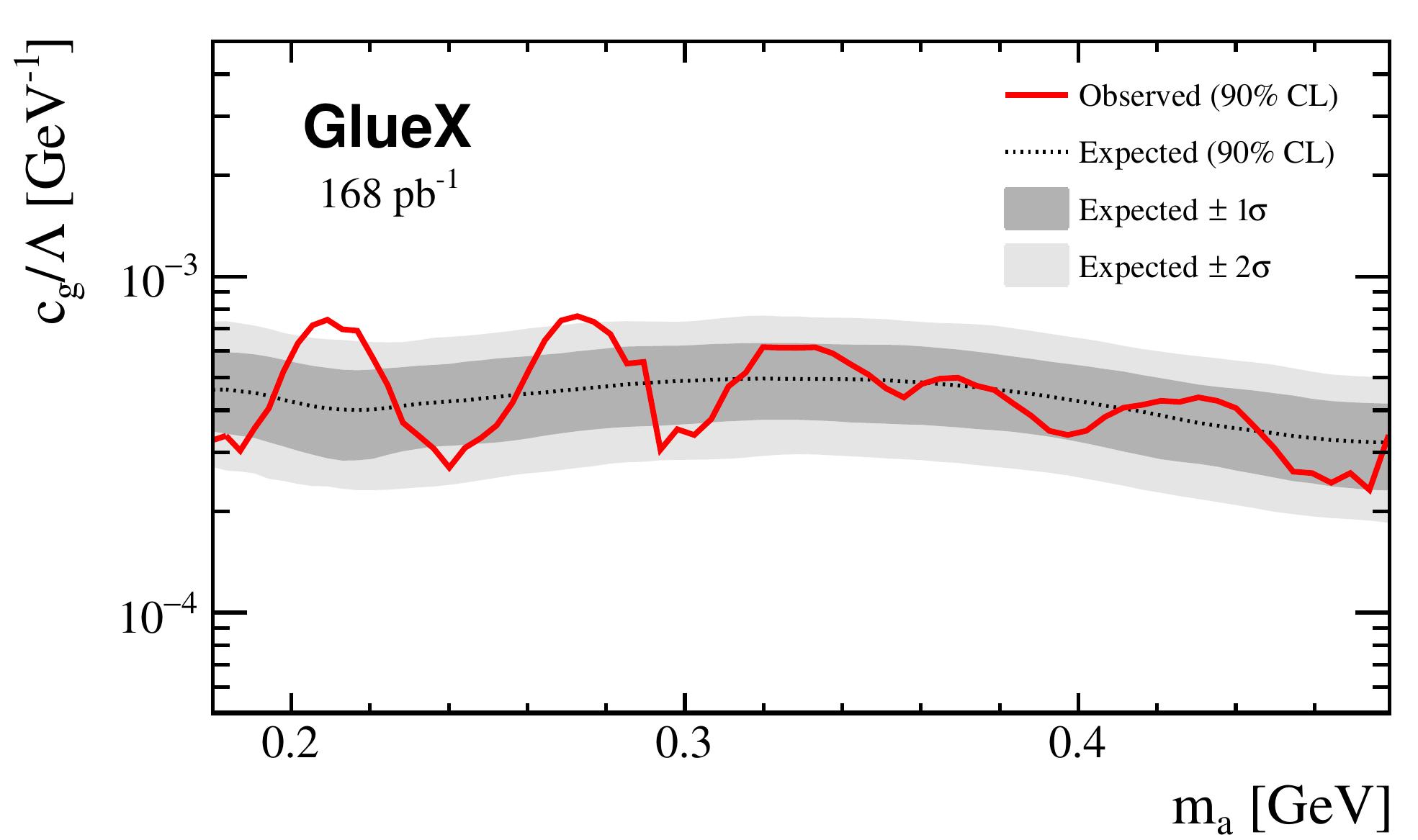}
  \includegraphics[width=0.45\textwidth]{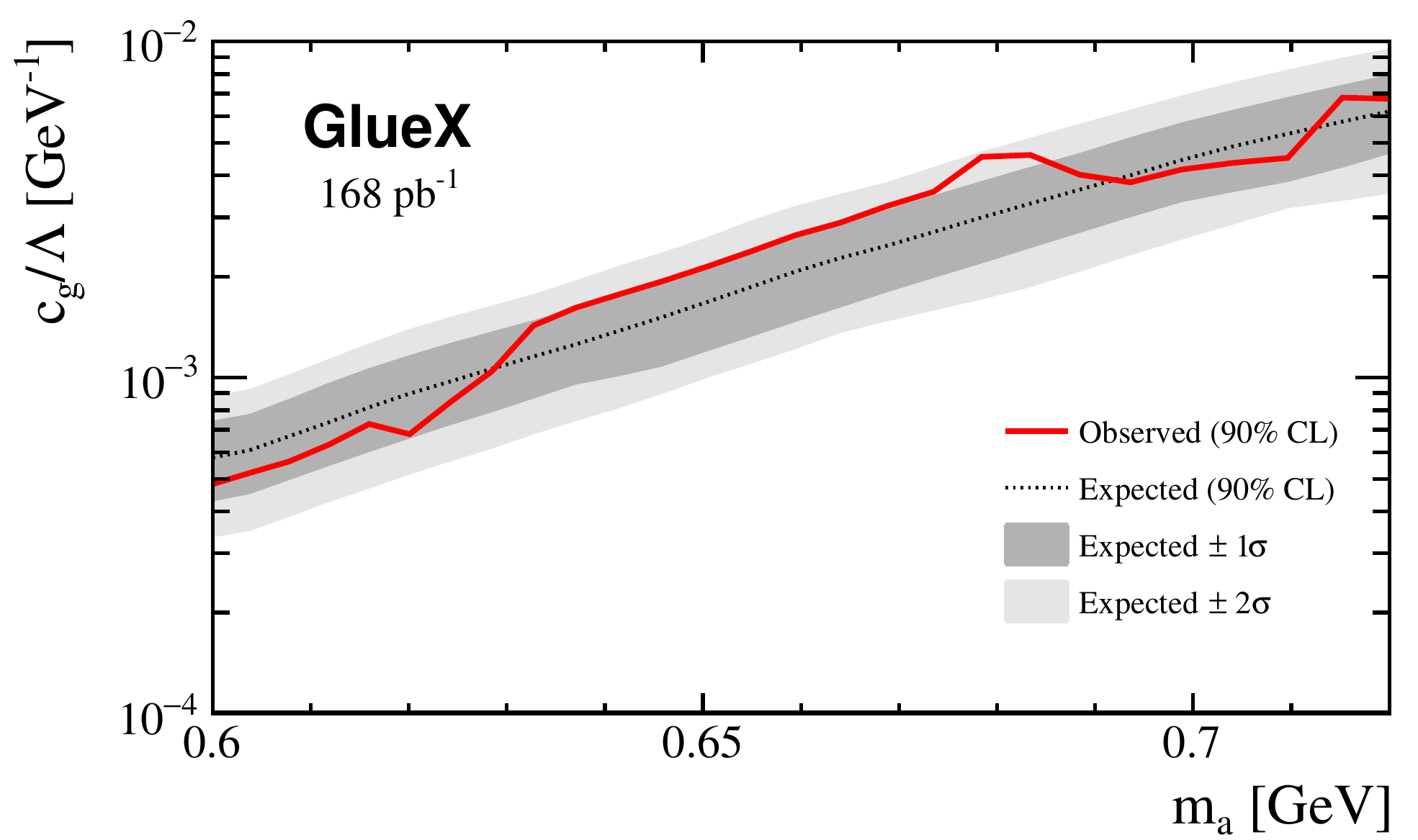}
  \caption{Exclusion limits at 90\% CL (red solid) from this search compared to the expected sensitivity (dashed) from (top) \diphoton and (bottom) \threepi channels. The (dark shaded) $\pm 1\sigma$ and (light shaded) $\pm 2\sigma$ regions are also shown.
  }
  \label{fig:limits}
\end{figure}

ALPs are excluded at 90\% confidence level (CL) when the upper limit on the observed ALP yield is less than the expected ALP yield in Eq.~\ref{eq:norm}.
Figure~\ref{fig:limits} shows the constraints placed on the ratio of ALP parameters $c_g / \Lambda$ for each final state.
Taking $c_g$ to be $\mathcal{O}(1)$, our results correspond to $\mathcal{O}({\rm TeV})$ constraints on $\Lambda$.

\begin{figure*}[t]
  \centering
  \includegraphics[width=0.7\textwidth]{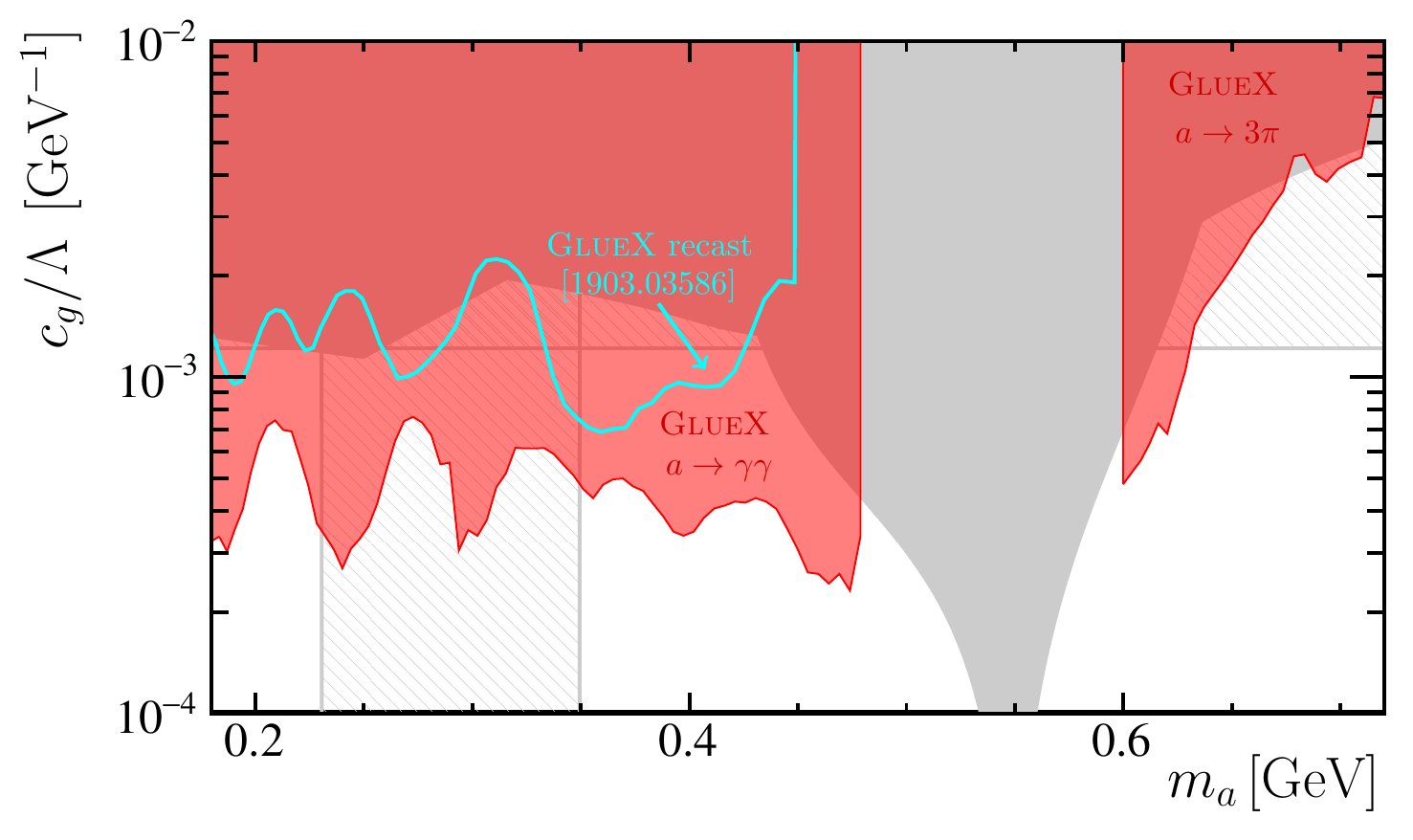}
  \caption{Results from (red) this search compared to the (gray) previous bounds~\cite{Williams_ALP_pheno}
  from LEP~\cite{LEP_PRL,LEP_OPAL}, $\phi$ and $\eta'$ decays~\cite{PDG}, and the (cyan line) \gluex limits recast from Ref.~\cite{GlueX_2017}.
  In addition, limits obtained from kaon decays~\cite{kaon1,kaon2,kaon3,Williams_ALP_pheno,Chakraborty:2021wda} and the $B$-meson lifetime~\cite{Williams_ALP_pheno,Chakraborty:2021wda}, which  have $\mathcal{O}(1)$ uncertainties induced by the unknown UV physics, are shown as hashed regions.
  }
  \label{fig:exclusion}
\end{figure*}

Figure~\ref{fig:exclusion} compares our results to the best existing constraints on the ALP-gluon coupling.
The kaon-decay and $B$-lifetime constraints involve  penguin decays that proceed via loops that are sensitive to the unknown UV completion of the full theory. The constraints shown in Fig.~\ref{fig:exclusion} are taken from Refs.~\cite{Williams_ALP_pheno,Chakraborty:2021wda} which assume an $\mathcal{O}({\rm TeV})$ UV scale, though both references note that these constraints have $\mathcal{O}(1)$ uncertainties induced by the unknown UV physics.
The searches presented here place more robust limits---which are also the most stringent constraints over much of the mass ranges considered.
These results demonstrate the power of using high-intensity photon beams to search for low-mass physics beyond the Standard Model.

\section{Summary}

In summary,
we present a search for axion-like particles produced in photon-proton collisions at a center-of-mass energy of approximately 4\,GeV.
The search looked for $a\to\gamma\gamma$ and $a\to\pi^+\pi^-\pi^0$ decays in a data sample corresponding to an integrated luminosity of 168 pb$^{-1}$ collected with the \gluex detector. The search for $a\to\gamma\gamma$ decays was performed in the mass range of $180 < m_a < 480$\,MeV, while the search for $a\to\pi^+\pi^-\pi^0$ decays
explored the $600 < m_a < 720$\,MeV region.
No evidence for an ALP signal was found,
leading to 90\% confidence-level exclusion limits on the ALP-gluon coupling strength.
These constraints are the most stringent to date over much of the mass ranges considered.

\begin{acknowledgments}
We would like to acknowledge the outstanding efforts of the staff of the Accelerator and the Physics Division at Jefferson Lab that made the experiment possible.
This work was supported in part by the U.S.\ Department of Energy, the U.S.\ National Science Foundation, the German Research Foundation, GSI Helmholtzzentrum f\"ur Schwerionenforschung GmbH, the Natural Sciences and Engineering Research Council of Canada, the Russian Foundation for Basic Research, the UK Science and Technology Facilities Council, the Chilean Comisi\'on Nacional de Investigaci\'on Cient\'ifica y Tecnol\'ogica, the National Natural Science Foundation of China and the China Scholarship Council. This material is based upon work supported by the U.S.\ Department of Energy, Office of Science, Office of Nuclear Physics under contract DE-AC05-06OR23177.
\end{acknowledgments}

\bibliographystyle{h-physrev}
\bibliography{refs}

\begin{thebibliography}{10}

\bibitem{Essig_2013}
R.~Essig {\em et~al.},
\newblock Dark sectors and new, light, weakly-coupled particles, 2013,
  1311.0029.

\bibitem{Marsh_2016}
D.~J.~E. Marsh,
\newblock Phys. Rep. {\bf 643}, 1 (2016).

\bibitem{Graham_experiment_2015}
P.~W. Graham, I.~G. Irastorza, S.~K. Lamoreaux, A.~Lindner, and K.~A. van
  Bibber,
\newblock Annu. Rev. Nucl. Part. Sci. {\bf 65}, 485 (2015).

\bibitem{Irastorza_2018}
I.~G. Irastorza and J.~Redondo,
\newblock Prog. Part. Nucl. Phys. {\bf 102}, 89 (2018).

\bibitem{Peccei_cp_1977}
R.~D. Peccei and H.~R. Quinn,
\newblock Phys. Rev. Lett. {\bf 38}, 1440 (1977).

\bibitem{Peccei_constraints_1977}
R.~D. Peccei and H.~R. Quinn,
\newblock Phys. Rev. D {\bf 16}, 1791 (1977).

\bibitem{Weinberg_1978}
S.~Weinberg,
\newblock Phys. Rev. Lett. {\bf 40}, 223 (1978).

\bibitem{Wilczek_1978}
F.~Wilczek,
\newblock Phys. Rev. Lett. {\bf 40}, 279 (1978).

\bibitem{Graham_hierarchy_2015}
P.~W. Graham, D.~E. Kaplan, and S.~Rajendran,
\newblock Phys. Rev. Lett. {\bf 115}, 221801 (2015).

\bibitem{Nomura_portal_2009}
Y.~Nomura and J.~Thaler,
\newblock Phys. Rev. D {\bf 79}, 075008 (2009).

\bibitem{Freytsis:2010ne}
M.~Freytsis and Z.~Ligeti,
\newblock Phys. Rev. {\bf D83}, 115009 (2011), 1012.5317.

\bibitem{Dolan:2014ska}
M.~J. Dolan, F.~Kahlhoefer, C.~McCabe, and K.~Schmidt-Hoberg,
\newblock JHEP {\bf 03}, 171 (2015), 1412.5174,
\newblock [Erratum: JHEP07,103(2015)].

\bibitem{Hochberg:2018rjs}
Y.~Hochberg, E.~Kuflik, R.~Mcgehee, H.~Murayama, and K.~Schutz,
\newblock (2018), 1806.10139.

\bibitem{Marciano:2016yhf}
W.~J. Marciano, A.~Masiero, P.~Paradisi, and M.~Passera,
\newblock Phys. Rev. {\bf D94}, 115033 (2016), 1607.01022.

\bibitem{Jaeckel:2015jla}
J.~Jaeckel and M.~Spannowsky,
\newblock Phys. Lett. {\bf B753}, 482 (2016), 1509.00476.

\bibitem{Dobrich:2015jyk}
B.~Dobrich, J.~Jaeckel, F.~Kahlhoefer, A.~Ringwald, and K.~Schmidt-Hoberg,
\newblock JHEP {\bf 02}, 018 (2016), 1512.03069,
\newblock [JHEP02,018(2016)].

\bibitem{Izaguirre:2016dfi}
E.~Izaguirre, T.~Lin, and B.~Shuve,
\newblock Phys. Rev. Lett. {\bf 118}, 111802 (2017), 1611.09355.

\bibitem{Knapen:2016moh}
S.~Knapen, T.~Lin, H.~K. Lou, and T.~Melia,
\newblock Phys. Rev. Lett. {\bf 118}, 171801 (2017), 1607.06083.

\bibitem{Williams_ALP_pheno}
D.~Aloni, Y.~Soreq, and M.~Williams,
\newblock Phys. Rev. Lett. {\bf 123}, 031803 (2019).

\bibitem{Williams_ALP_photoproduction}
D.~Aloni, C.~Fanelli, Y.~Soreq, and M.~Williams,
\newblock Phys. Rev. Lett. {\bf 123}, 071801 (2019).

\bibitem{Kelly:2020dda}
K.~J. Kelly, S.~Kumar, and Z.~Liu,
\newblock (2020), 2011.05995.

\bibitem{Hook:2014cda}
A.~Hook,
\newblock Phys. Rev. Lett. {\bf 114}, 141801 (2015), 1411.3325.

\bibitem{Fukuda:2015ana}
H.~Fukuda, K.~Harigaya, M.~Ibe, and T.~T. Yanagida,
\newblock Phys. Rev. D {\bf 92}, 015021 (2015), 1504.06084.

\bibitem{Dimopoulos:2016lvn}
S.~Dimopoulos, A.~Hook, J.~Huang, and G.~Marques-Tavares,
\newblock JHEP {\bf 11}, 052 (2016), 1606.03097.

\bibitem{Hook:2019qoh}
A.~Hook, S.~Kumar, Z.~Liu, and R.~Sundrum,
\newblock Phys. Rev. Lett. {\bf 124}, 221801 (2020), 1911.12364.

\bibitem{Kamionkowski:1992mf}
M.~Kamionkowski and J.~March-Russell,
\newblock Phys. Lett. B {\bf 282}, 137 (1992), hep-th/9202003.

\bibitem{Barr:1992qq}
S.~M. Barr and D.~Seckel,
\newblock Phys. Rev. D {\bf 46}, 539 (1992).

\bibitem{Ghigna:1992iv}
S.~Ghigna, M.~Lusignoli, and M.~Roncadelli,
\newblock Phys. Lett. B {\bf 283}, 278 (1992).

\bibitem{Holman:1992us}
R.~Holman {\em et~al.},
\newblock Phys. Lett. B {\bf 282}, 132 (1992), hep-ph/9203206.

\bibitem{GlueX_2017}
GlueX Collaboration, H.~Al~Ghoul {\em et~al.},
\newblock Phys. Rev. C {\bf 95}, 042201 (2017).

\bibitem{PDG}
Particle Data Group, M.~Tanabashi {\em et~al.},
\newblock Phys. Rev. D {\bf 98}, 030001 (2018).

\bibitem{Supp}
See Supplemental Material at the end of this article for details and numerical
  results required to recast these results for other models.

\bibitem{CEBAF2001}
C.~W. Leemann, D.~R. Douglas, and G.~A. Krafft,
\newblock Annual Review of Nuclear and Particle Science {\bf 51}, 413 (2001).

\bibitem{GlueX_NIM}
S.~Adhikari {\em et~al.},
\newblock Nucl. Instrum. and Methods Sect. A {\bf 987}, 164807 (2021).

\bibitem{SC_NIM}
E.~Pooser {\em et~al.},
\newblock Nucl. Instrum. and Methods Sect. A {\bf 927}, 330 (2019).

\bibitem{CDC_NIM}
N.~Jarvis {\em et~al.},
\newblock Nucl. Instrum. and Methods Sect. A {\bf 962}, 163727 (2020).

\bibitem{BCAL_NIM}
T.~Beattie {\em et~al.},
\newblock Nucl. Instrum. and Methods Sect. A {\bf 896}, 24 (2018).

\bibitem{FDC_NIM}
L.~Pentchev {\em et~al.},
\newblock Nucl. Instrum. and Methods Sect. A {\bf 845}, 281 (2017),
\newblock Proceedings of the Vienna Conference on Instrumentation 2016.

\bibitem{FCAL_NIM}
K.~Moriya {\em et~al.},
\newblock Nucl. Instrum. and Methods Sect. A {\bf 726}, 60 (2013).

\bibitem{genr8}
{\sc Genr8} Monte Carlo Event Generator, GlueX-doc-11 (1998).

\bibitem{geant4}
S.~Agostinelli {\em et~al.},
\newblock Nucl. Instrum. and Methods Sect. A {\bf 506}, 250 (2003).

\bibitem{Skwarnicki:1986xj}
T.~Skwarnicki,
\newblock {\em {A study of the radiative cascade transitions between the
  Upsilon-prime and Upsilon resonances}},
\newblock PhD thesis, Institute of Nuclear Physics, Krakow, 1986,
\newblock {\href{http://inspirehep.net/record/230779/}{DESY-F31-86-02}}.

\bibitem{omega_lineshape}
M.~N. Achasov {\em et~al.},
\newblock Phys. Rev. D {\bf 68}, 052006 (2003).

\bibitem{Williams_bump_hunt}
M.~Williams,
\newblock Journal of Instrumentation {\bf 12}, P09034 (2017).

\bibitem{LHCb_dark_photon_2018}
LHCb Collaboration, R.~Aaij {\em et~al.},
\newblock Phys. Rev. Lett. {\bf 120}, 061801 (2018).

\bibitem{LHCb_dark_photon_2020}
LHCb Collaboration, R.~Aaij {\em et~al.},
\newblock Phys. Rev. Lett. {\bf 124}, 041801 (2020).

\bibitem{Aaij:2020ikh}
LHCb Collaboration, R.~Aaij {\em et~al.},
\newblock JHEP {\bf 10}, 156 (2020).

\bibitem{Rolke:2004mj}
W.~A. Rolke, A.~M. Lopez, and J.~Conrad,
\newblock Nucl. Instrum. Meth. {\bf A551}, 493 (2005), physics/0403059.

\bibitem{Dauncey:2014xga}
P.~D. Dauncey, M.~Kenzie, N.~Wardle, and G.~J. Davies,
\newblock JINST {\bf 10}, P04015 (2015), 1408.6865.

\bibitem{Chakraborty:2021wda}
S.~Chakraborty, M.~Kraus, V.~Loladze, T.~Okui, and K.~Tobioka,
\newblock (2021), 2102.04474.

\bibitem{LEP_PRL}
S.~Knapen, T.~Lin, H.~K. Lou, and T.~Melia,
\newblock Phys. Rev. Lett. {\bf 118}, 171801 (2017).

\bibitem{LEP_OPAL}
OPAL Collaboration, G.~Abbiendi~et al. {\em et~al.},
\newblock Eur. Phys. J. C {\bf 26}, 331 (2003).

\bibitem{kaon1}
E.~Izaguirre, T.~Lin, and B.~Shuve,
\newblock Phys. Rev. Lett. {\bf 118}, 111802 (2017).

\bibitem{kaon2}
C.~Lazzeroni {\em et~al.},
\newblock Physics Letters B {\bf 732}, 65 (2014).

\bibitem{kaon3}
E.~Abouzaid {\em et~al.},
\newblock Phys. Rev. D {\bf 77}, 112004 (2008).

\end{thebibliography}

\end{document}